# GMISeg: General Medical Image Segmentation without Re-Training

Jing Xu[1, *]

1. Department of Computing, Imperial College London, London, UK

* j.xu23@imperial.ac.uk

**Abstract**: Deep learning models have become the dominant method for medical image segmentation. However, they often struggle to be generalisable to unknown tasks involving new anatomical structures, labels, or shapes. In these cases, the model needs to be re-trained for the new tasks, posing a significant challenge for non-machine learning experts and requiring a considerable time investment. Here I developed a general model that can solve unknown medical image segmentation tasks without requiring additional training. Given an example set of images and visual prompts for defining new segmentation tasks, GMISeg (General Medical Image Segmentation) leverages a pre-trained image encoder based on ViT and applies a low-rank fine-tuning strategy to the prompt encoder and mask decoder to fine-tune the model without in an efficient manner. I evaluated the performance of the proposed method on medical image datasets with different imaging modalities and anatomical structures. The proposed method facilitated the deployment of pre-trained AI models to new segmentation works in a user-friendly way.

**Keywords:** Model Generalisation, Low-rank Adaptation, Medical Image Segmentation.

## 1 Introduction

Image segmentation is an important yet classic topic in computer vision, particularly in the domain of analysing medical images. Managing tasks related to medical image segmentation entails navigating through various imaging modalities, like magnetic resonance imaging (MRI), X-ray, computed tomography (CT), and microscopy. These images may encompass various anatomical structures, such as the abdomen, chest, brain, retina, or individual cells. This variety in terms of both imaging modalities and anatomical structures has prompted the creation of numerous task-specific segmentation models, each crafted for a specific task for a particular kind of medical images [1]. Notably, in recent times, deep learning models have surfaced as the primary approach for medical image segmentation [2].

There are several major challenges in medical image segmentation and one of them is the domain shift challenge, which refers to the performance drop of a trained model on test data beyond the training data distribution. In the medical domain, this challenge is intensified, as clinical researchers and scientists continuously define new segmentation tasks influenced by evolving populations, scientific progress, and clinical objectives. To address this challenge, the image segmentation model typically requires training from scratch or fine-tuning using the new segmentation datasets. Training of a new segmentation model requires expertise in machine learning, substantial computational resources, and human effort—often exceeding the reach of clinical researchers or scientists lacking the necessary knowledge and tools for model training. This practical limitation impedes the pace of developing and deploying segmentation models in clinical research. Therefore, the primary research aim of this thesis involves addressing new segmentation tasks without the necessity for model re-training.

Finetuning models pre-trained on natural images proves unproductive in the medical domain [3]. Discrepancies in image features, and task specifications across domains may contribute to this inefficacy. Importantly, a substantial amount of retraining still necessary. While few-shot semantic segmentation methods aim to predict new classes within a limited data regimes without fine-tuning, they are mainly designed for classification or segmentation tasks within the same domain [4]. These methods struggle to generalize across imaging modalities or anatomical structures [3].

This thesis aims to develop a versatile medical image segmentation method, GMISeg, designed to generalise well in diverse tasks without the need for retraining, e. GMISeg achieves this by learning how to use the input of a set of labeled samples from a specified segmentation taskto perform segmentation on new biomedical images through forward transfer. The key contributions are outlined as follows:

(1) I proposed GMISeg, a framework that can perform new segmentation tasks without further training. GMISeg uses a new low-rank fine-tuning strategy, which improves the model's anti-forgetting ability and only adds and updates a small number of parameters during the fine-tuning process.

(2) I evaluated the performance of GMISeg and compare it to state-of-the-art segmentation models in tasks involving different imaging modalities and anatomical structures.

(3) I explored the performance of the proposed framework for diverse tasks, and the influence of the support set size on the generalisation ability of the algorithm.

## 2 Related works

This section will briefly introduce the terminology, methods, and progress of medical image segmentation. Then, it will outline the basic methods for adapting models to small sample data and unseen data, including multi-task learning, transfer learning, optimization-based meta-learning, and few-shot semantic segmentation.

**Medical Image Segmentation.** In the realm of medical image segmentation, in-depth research entails training convolutional neural networks (CNN) in a supervised manner to forecast label mappings for input images [5]. Typically, addressing new segmentation challenges involves training models from the ground up, demanding substantial design efforts and adjustments.

Unlike recent models like nnUNet [6], which automate specific design decisions such as data processing or model architecture but still incur significant training overhead, GMISeg distinguishes itself by enabling generalization to new medical image segmentation tasks without requiring additional training.

**Multi-task Learning.** The multi-task learning (MTL) framework involves simultaneously learning multiple tasks [7], which, in the context of medical imaging, can encompass various forms [8], population centers [9], or anatomical sites [10]. However, traditional MTL approaches predetermine tasks during the design phase, restricting networks to solving only those tasks identified during training. GMISeg breaks free from this constraint by enabling dynamic specification of tasks during the inference process.

**Transfer Learning.** Transfer learning strategies frequently involve fine-tuning pre-trained models, often sourced from different fields [11]. In the context of medical image segmentation, this includes starting with models trained on natural images [12], leveraging the substantial data available in this domain. Despite its commonality, this approach requires extensive training for each new task, a challenge overcome by GMISeg. Importantly, the differences between medical images and natural images often render transfer learning from large pre-trained models less effective [3].

**Optimization-based Meta-Learning.** Optimization-based meta-learning techniques, commonly minimizing downstream fine-tuning steps through multiple examples per task, known as few-shot learning, have been explored in medical image segmentation [13].

Meta-learning via fine-tuning has been studied in medical image segmentation, including handling multiple image modalities[14], anatomy [15], and generalizing to different targets domains [16]. While these strategies mitigate data and training demands for downstream tasks [17], it is crucial to emphasize that fine-tuning such models still requires machine learning expertise and computational resources, often presenting challenges for medical researchers.

**Few-shot Semantic Segmentation.** The Few-Shot (FS) method typically adapts to new tasks using a small number of training examples by fine-tuning a pre-trained network [18]. Some few-shot semantic segmentation models can perform predictions for new images with unseen classes, relying on only a few labeled examples without requiring additional retraining. A prevalent strategy in FS segmentation, applicable to both natural images [19] and medical images [20], involves leveraging large pre-trained models to extract deep features from query and support images. These methods commonly focus on learning meaningful prototype representations for each label [21]. Another strategy in medical FS segmentation employs self-supervised learning to address training data and task limitations [22]. Unlike GMISeg, these methods concentrate on improving the segmentation performance within the limited data scenarios for the same segmentation task, such as abdominal CT or MRI segmentation [23, 24], and are only suitable for low-dimensional and tabular data [25]. Our solution is inspired by the ideas of these methods but aims to solve the segmentation task of different images, such as multi-modal medical images of different anatomical structures.

Although there are currently some fine-tuning models based on SAM, however mask decoders have also been fine-tuning [37] and used for image classification [38].

In this research, the commitment is to abstain from any re-training even when ample examples are provided for a new task. This approach is designed to spare clinical or scientific users from the necessity of possessing machine learning expertise and substantial computing resources. This work broaden the model's applicability to a wide range of anatomical organs and imaging modalities, including those entirely unseen during the training process.

## 3 GMISeg

Given a medical image $I$ with a spatial resolution of $H \times W$ and a channel number of $x \in R^{H \times W \times C}$ $I$, the method aims to predict its corresponding segmentation $\hat{S}$ at the resolution of $H \times W$, where each pixel belongs to an element in the predefined class list $Y = \{y_0, y_1, \cdots, y_k\}$, as close as possible to the real segmentation $S$. $y_0$ denotes the background class and $y_i, i \in \{1, \cdots, k\}$ denotes the class for different organs. As shown in Figure 1, the overall architecture of GMISeg is inherited from SAM [26]. This method freezes all parameters in the image encoder and designs a trainable bypass for each transformer block. As shown in Loose LoRA (Loose Low-Rank Adaptation), these bypasses first compress the transformed features into low-level space, and then re-project the compressed features to align with the channels of the output features in the frozen transformer block.

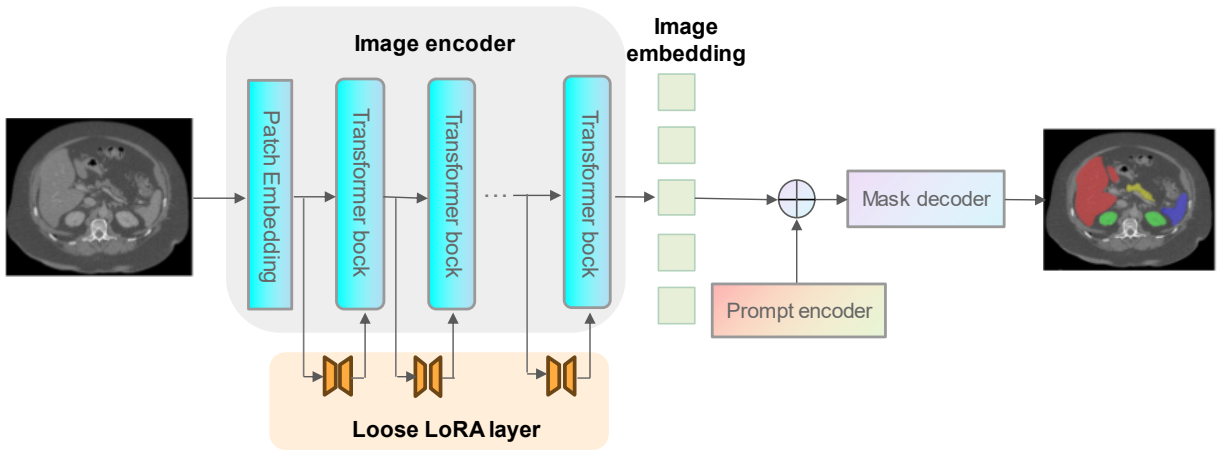

Fig. 1: **The structure of GMISeg.** The framework of GMISeg is consistent with SAM. This method freeze the image encoder and inserted the proposed trainable Loose LORA layer into SAM for image feature extraction.

### 3.1 LoRA in Image Encoder

The image encoder of GMISeg mainly consists of modules based on ViT (Vision Transformer) [27], However, when fine-tuning a pretrained ViT model for a new task, a key issue is ensuring that the model does not forget its original knowledge [28]. To address this, this study plan to propose a novel approach, applying the LoRA structure to the Q, K, V projection layers of ViT, further enhancing its expressiveness and robustness.

In the standard ViT framework, V, K, and Q and stand for Value, Key, and Query respectively, which form the core components of the self-attention mechanism in the Transformer. As illustrated in Figure 2, we introduce LoRA layers to Q and V, enabling them to undergo parameter updates more efficiently while retaining the original knowledge. Specifically, this design allows the model to update only a small portion of its parameters during fine-tuning rather than fully retraining the entire model.

LoRA essentially performs low-rank approximation, which can be viewed as a specific parameterisation method capable of approximating a large transformation matrix with a limited number of parameters from low-rank matrices [29]. The motivation for this design is that when fine-tuning a model, task-related information often resides in a low-dimensional subspace rather than being scattered across the entire parameter space. Hence, with LoRA, the model can solely update this low-dimensional subspace, leading to more efficient fine-tuning.

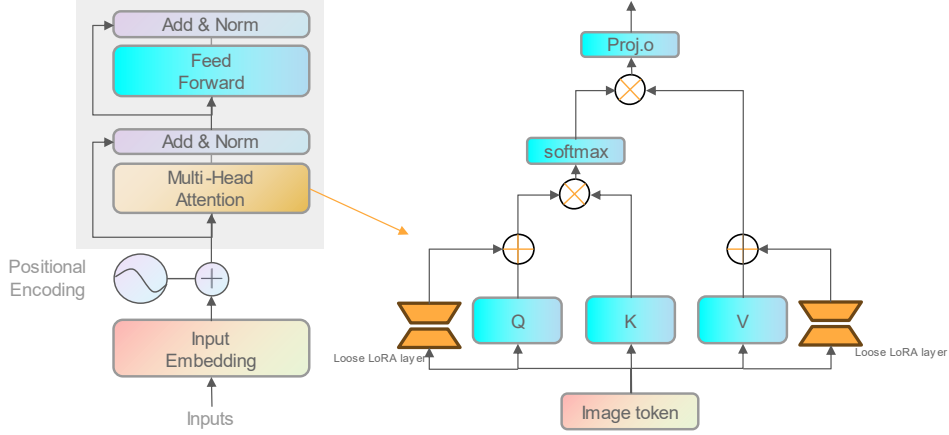

Fig. 2: **The structure of Loose LoRA in image encoder.** This study applies the Loose LoRA layer to the $q$ and $v$ projection layers of each transform block in the image encoder. '$Q$', '$K$', '$V$', and '$Proj.o$' represent the projection layers of $q$, $k$, $v$, and $o$.

Equation (1) described how the weight matrix $W$ is updated, where $A$ and $B$ are two linear layers, $\triangle W$ is the cumulative gradient update during the finetuning process, and $F$ is the sequence of input tokens.

$$\hat{F} = WF$$
$$\hat{W} = W + \triangle W = W + BA \tag{1}$$

In this design, LoRA not only encompasses the MLP structure of the encoder and decoder but also integrates an average pooling structure. The inclusion of average pooling enhances the rigidity of the model, allowing it to better retain original knowledge. Simultaneously, it boosts the plasticity of the model, enabling it to adapt better to new tasks.

Equation (2) explains how $Q, K, V$ are computed, where $W_q$ and $W_k$ are pretrained projection layers, and $A_q$ and $B_q$ are trainable LoRA parameters, $T$ is the transformer's multi headed attention, and $C_{out}$ is variance.

$$tt(Q, K, V) = Softmax\left(\frac{Q^T K}{\sqrt{C_{out}}} + B\right) V$$

$$Q = \widehat{W_q}F = W_q F + B_q A_q F \tag{2}$$

$$K = W_k F$$

### 3.2 Loose LoRA Embedding

Usually, the incorporated LoRA layer is the MLP structure in the encoder and decoder. However, there is a problem with this structure, which may cause catastrophic forgetting for large models. This means when a new dataset is used for model finetuning, the model will forget the pre-trained knowledge. Here we propose a loose LoRA layer, that is, each MLP (each layer of MLP, including the encoder and decoder) is transformed into a vector of the same size through an average pooling operation. This structure is loose enough to improve the rigidity (anti-forgetting ability) and plasticity (learning ability and generalization ability) of the model.

Figure 3 shows that the Loose LoRA contains two parts, encoder and decoder, each encoder and decoder layers in Lora has different size. We extract average pooled features from each layer of the encoder and the decoder, and concatenate them to form a large matrix, which replaces the original LoRA layer.

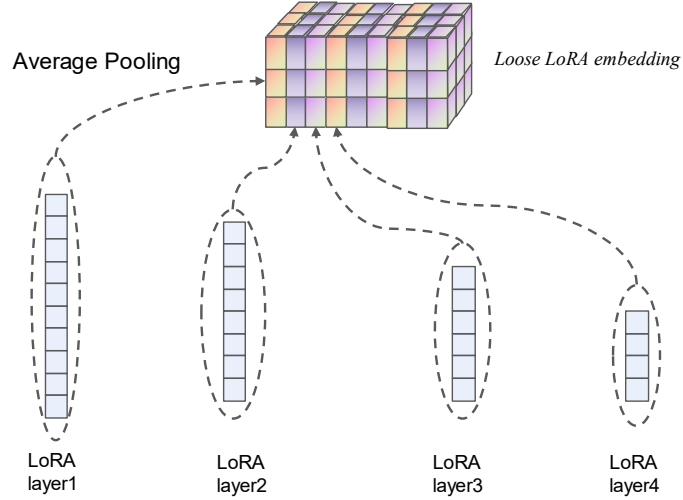

Fig. 3: **The structure of Loose LoRA embedding.** New LoRA converts each MLP (each layer of MLP, including encoder and decoder) into a vector of the same size through an average pooling operation. LoRA layers 1-4 mean 4 layers extracted from Lora's encoder and decoder.

### 3.3 Training strategies

**Loss Function**. GMISeg uses a combination of the focal loss and the Dice loss to perform fine-tuning, formulated as:

$$L = \mu_1 Focal\left(\widehat{S}_l, D(S)\right) + \mu_2 Dice\left(\widehat{S}_l, D(S)\right) \tag{3}$$

where $Focal$ and $Dice$ denote focal loss and Dice loss, respectively. $D$ represents the downsampling operation so that the resolution of ground truth is the same as that of GMISeg's output, because its spatial resolution is much lower. $\mu_1$ and $\mu_2$ are the weights to balance the impact between the two loss items.

**Prompt Encoder Generation**. Randomly select one or more points in GroundTruth to form a set of points, and select the maximum edge in GroundTruth to add an offset to generate a box, thus simulating user input.

### 4 Experiments

We first outline the datasets and implementation of the method. Subsequently, we plan to explore diverse analyses, encompassing the influence of training task diversity, the number of available examples for new tasks, and the size of the support set.

### 4.1 Datasets

The proposed model is trained and evaluated for a diverse set of segmentation tasks, using public medical segmentation challenge datasets with different anatomical structures, imaging modalities, and labels. The training set contains only the cardiac training image. The validation set contains only the cardiac validation image. The testing set consists of 5 datasets, covering 4 medical fields and 3 imaging modalities.

This study Standard the data in various formats of the original dataset, processed images, and label maps. Additionally, augmentation segmentation tasks incorporated to augment the training data, further enhancing the diversity of training tasks.

The dataset has a wide range of biomedical fields, such as the cardiac (ACDC [30], MICCAI 2018 [31]), abdomen (Synapses [1]), brain (ISLES 2022 [32]), and retina (STARE [33]). A detailed list of datasets is provided in Table 1. Despite the datasets cover various imaging tasks and labeling protocols, I focus in this work centres on the overarching challenge of 2D binary segmentation. In cases where the dataset includes 3D data, we extract the 2D middle slices of the 3D volume along XY-, XZ- and YZ- imaging planes.

Table 1: Test dataset.

| Dataset Name | Description | # of Scans | Image Modalities |
|---|---|---|---|
| ACDC | Left and right ventricular endocardium | 99 | Cine-MRI |
| MICCAI 2018 | Left atria | 100 | MRI |
| Synapes | Abdominal organs | 3779 | CT |
| STARE | Blood vessels in retinal images | 20 | Optical camera |
| ISLES 2022 | Ischemic stroke lesion | 180 | Multi-modal MRI |

### 4.2 Implementation details

This study adopted the data enhancement strategy of random flip and random gaussian noise and carried out all experiments based on the "vit_L" version of SAM. This study first upsample all images to $1024 \times 1024$, and then input the up-sampled image into GMISeg to maintain a better image resolution of the predicted label probability map. The Loose LoRA is used to fine-tune the frozen $Q$ and $V$ projection layers of the transformer block. The rank of LoRA is set to 4 for efficiency and performance optimization. The weight of focal loss set is 1, and the weight of Dice loss set is 1.

### 4.3 Task Generalization Results

This study conducted a segmentation performance comparison, pitting GMISeg against the latest generalized segmentation models, namely CycleGAN [34], Neuralizer [35], and Universeg [36]. The primary goal was to evaluate the effectiveness of GMISeg in handling tasks from unknown datasets. Table 2 outlines the average Dice score, Jaccard score, and ASD values for each dataset and method, while Figure 4 provides a visual representation of segmentation results for each dataset and method.

Table 2: **Performance summary.** This study reports the average Dice score, Jaccard score, and ASD values for each dataset of each model.

| Method | Dataset | Dice (%) | Jaccard (%) | ASD (mm) |
|---|---|---|---|---|
| **GMISeg** | ACDC | **78.7** | **74.5** | **1.2** |
| CycleGAN | ACDC | 75.5 | 72.1 | 2.4 |
| UniverSeg | ACDC | 75.0 | 71.6 | 2.7 |
| Neuralizer | ACDC | 74.7 | 71.2 | 3.1 |
| **GMISeg** | MICCAI 2018 | **77.8** | **84.3** | **1.3** |
| CycleGAN | MICCAI 2018 | 75.3 | 71.9 | 2.5 |
| UniverSeg | MICCAI 2018 | 74.9 | 71.4 | 2.8 |
| Neuralizer | MICCAI 2018 | 74.5 | 71.0 | 3.3 |
| **GMISeg** | Synapes | **77.5** | **73.9** | **1.4** |
| CycleGAN | Synapes | 75.0 | 71.5 | 2.6 |
| UniverSeg | Synapes | 74.6 | 71.1 | 3.0 |
| Neuralizer | Synapes | 74.2 | 70.8 | 3.4 |
| **GMISeg** | STARE | **77.6** | **74.0** | **1.5** |
| CycleGAN | STARE | 74.9 | 71.3 | 2.9 |
| UniverSeg | STARE | 74.4 | 70.9 | 3.2 |
| Neuralizer | STARE | 74.0 | 70.5 | 3.6 |
| **GMISeg** | ISLES 2022 | **78.3** | **76.3** | **1.6** |
| CycleGAN | ISLES 2022 | 74.3 | 69.9 | 3,6 |
| UniverSeg | ISLES 2022 | 74.9 | 73.2 | 3.0 |
| Neuralizer | ISLES 2022 | 74.5 | 72.8 | 2.9 |

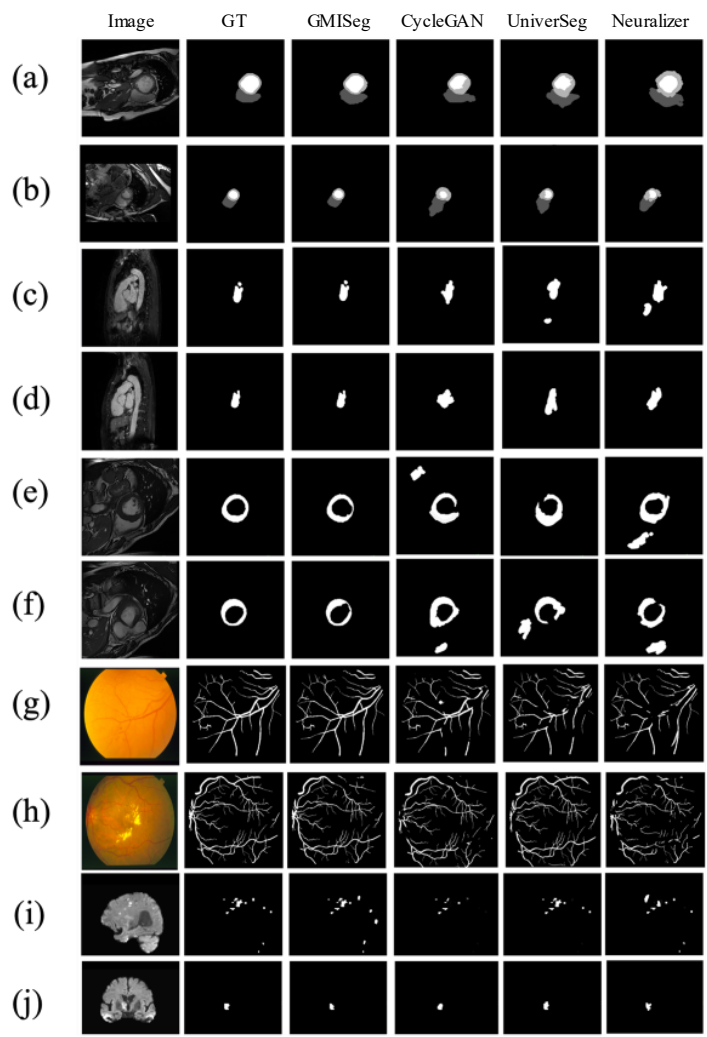

Fig. 4: **Example model predictions for unseen tasks.** For randomly sampled images of each segmentation task, this study visualized the predictions of GMISeg and the three latest models, as well as GroundTruth maps. Among them, (a) (b) comes from ACDC, (c) (d) comes from MICCAI 2018, (e) (f) comes from Synapse, (g) (h) comes from STARE, and (i) (j) comes from ISLES 2022.

Table 2 shows that in all datasets, GMISeg 's performance is significantly better than all the latest methods, with Dice score's improvement range from 3.2% to 4.0%, Jaccard's improvement range from 2.4% to 13.3%, and ASD's improvement range from 1.2mm to 2.1mm.

Figure 4 also shows the obvious quality improvement in generalization segmentation. Compared with other methods, the segmented region predicted by GMISeg is smoother and more accurate. This study attributes this phenomenon to the powerful feature extraction ability of large-scale SAM models and the effective fine-tuning strategy adopted by GMISeg.

### 4.4 Analysis

**Task Quantity and Diversity.** This study investigated the number of datasets used to train GMISeg and the impact of individual tasks. This study omitted the synthesis task in this experiment and trained the model on a random subset of the training dataset.

Figure 5 illustrates the performance of various random subsets of the training dataset on the test dataset. This work findings indicate that a greater number of training tasks correlates with improved performance on test tasks. Notably, the selection of datasets can significantly impact results. For instance, in a model trained with a 10% dataset, the best-performing model achieves a Dice score nearly 23% higher than the worst-performing model.

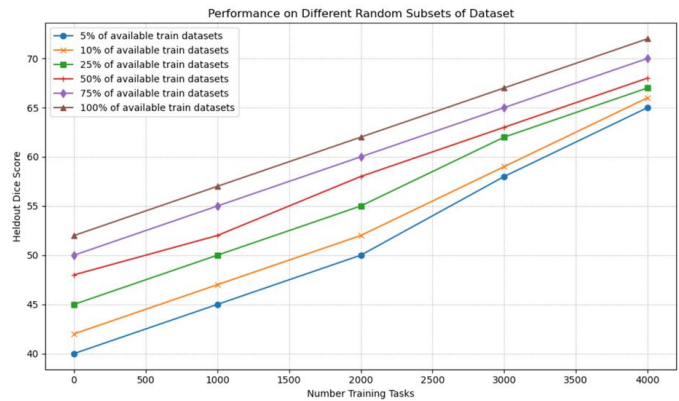

Fig. 5: The plot depicts the Average Dice score in relation to the number of training tasks. Each data point represents an individual network trained based on the percentage of available training datasets, and they are differentiated by the number of basic training tasks.

**Support Set Size.** This study explored the influence of support size on the training model, varying the support size (N) from 1 to 64.

As depicted in Figure 6, Optimal results are obtained with a larger training support size. The average Dice score exhibits a rapid increase from 57.6% to 69.1% as support sizes range from 1 to 64. Notably, this study observed that ensemble prediction contributes to further improvement when the support size exceeds 6 ($N > 6$).

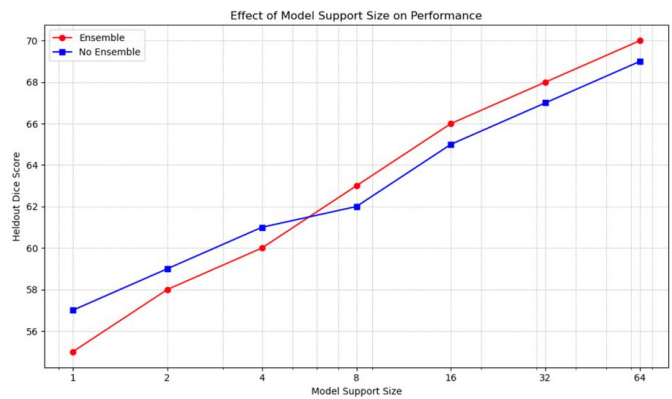

Fig. 6: The relationship between models trained at a specific support set size and their average Dice scores reveals that increasing the support size leads to improved results. Moreover, integration consistently proves beneficial across varying support sizes.

**Limited Example Data.** Given the high cost of manually annotating examples in new medical tasks, this study aimed to comprehend how the number of labeled images influences GMISeg's performance. This study conducted experiments using a restricted number of labeled examples (n) for reasoning, where $n = 1, 2, 64$. Each size underwent 100 replicates, each representing an independent random subset of the data. In this context, the support set includes all available data for reasoning, eliminating the need for integration.

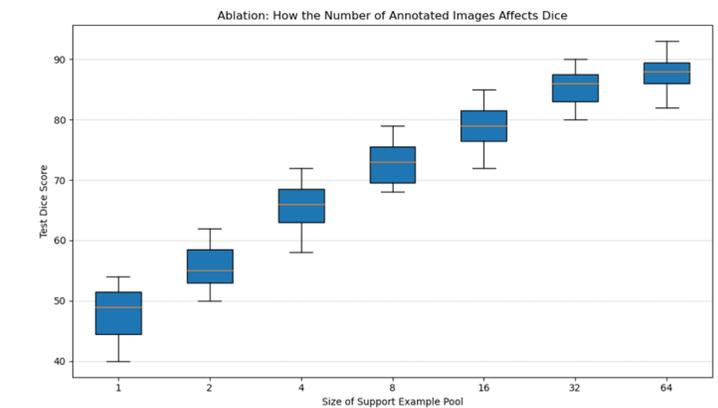

Fig. 7: This study explored the impact of available data on GMISeg predictions using limited data samples on the supporting STARE dataset. Conducting 100 replicates for each size, this study employed different random subsets to assess the consistency of results.

Figure 7 illustrates the results for the STARE dataset. Notably, for smaller support sizes (n), significant differences emerge due to the highly diverse support set. As n increases, a monotonic improvement in average segmentation quality is observed, accompanied by a notable decrease in variance with the available data sample.

## 5 Conclusion

GMISeg is introduced as an agnostic model designed for learning medical image segmentation. Leveraging a substantial and diverse set of public medical segmentation datasets, GMISeg is trained and tested with the capability to generalize to unknown anatomy and tasks. This study proposed a novel low-rank fine-tuning strategy, which improves the anti-forgetting ability of the model by extracting and merging the loose layer structure, and it can only add and update a small number of parameters in the customization process, thus improving the learning ability and generalization ability of model.

In the experiment, the results of GMISeg on all datasets are much better than the latest generalized medical image segmentation methods. Through extensive in-depth research, this study concluded that the performance of GMISeg strongly depends on task diversity during training, the number of available examples for new tasks, and the size of the support set during reasoning.

In this study, 2D data and a single tag were employed to emphasize and analyze the core functions of GMISeg. Future plans include extending the model to incorporate 3D models and multi-label map segmentation of 3D volumes. GMISeg is anticipated to adeptly handle new tasks defined by researchers and scientists without the impractical need for model retraining. This study envisions GMISeg garnering increased attention in the field of medical imaging and aim to develop a more effective general segmentation model.

### Acknowledgement

This study thanks NVIDIA Corporation for the GPU donation.